\documentclass[doublecol]{epl2}

\usepackage{amsmath}
\usepackage{amssymb}
\usepackage{empheq}

\pdfoutput=1

\title{Modulation instability and capillary wave turbulence}
\shorttitle{Modulation instability and capillary wave turbulence} 

\author{ H. Xia\thanks{E-mail: \email{hua.xia@anu.edu.au}} \and M. Shats \and H. Punzmann}
\shortauthor{H. Xia, M. Shats, H. Punzmann}

\institute{
Research School of Physics and Engineering, The Australian National University \\ Canberra, ACT 0200, Australia}

\pacs{47.35.Pq}{Capillary wave}
\pacs{05.45.-a}{nonlinear dynamics}

\abstract{Formation of turbulence of capillary waves is studied in laboratory experiments. The spectra show multiple exponentially decreasing harmonics of the parametrically excited wave which nonlinearly broaden with the increase in forcing. Spectral broadening leads to the development of the spectral continuum which scales as $\propto f^{-2.8}$, in agreement with the weak turbulence theory (WTT) prediction. Modulation instability of capillary waves is shown to be responsible for the transition from discrete to broadband spectrum. The instability leads to spectral broadening of the harmonics, randomization of their phases, it isolates the wave field from the wall, eventually allows the transition from 4- to 3-wave interactions as the dominant nonlinear process, thus creating the prerequisites assumed in WTT.}

\begin{document}

\maketitle

Broad frequency and wave number spectra characterize many turbulent systems. When energy is injected into a system in a limited range of wave numbers or frequencies, nonlinear interactions extend this range. In the flows described by the Navier-Stokes equations such spectral energy transfer at large Reynolds numbers is attributed to the energy cascades. The idea of the cascade also lies in the center of the weak wave turbulence theory (WTT) \cite{Zakharov_Lvov_Falkovich}, where instead of strong nonlinearity of the Navier-Stokes equation, the cascade is driven by weakly interacting waves having almost random phases. Among the best known applications of the WTT to the surface waves is the theory of capillary wave turbulence of Zakharov and Filonenko \cite{Zakharov_1967}.

The theory assumes that the dispersion relation of capillary waves $\omega_k=(\alpha k^3/\rho)^{1/2}$ (where $\alpha$ is the surface tension and $\rho$ is the liquid density) allows simultaneous satisfaction of the matching rules for the wave triads in the frequency and in the wave number domains "for any possible choice of the wave vectors" $\vec{k}_1$,$\vec{k}_2$,$\vec{k}_3$ \cite{Pushkarev&Zakharov2000}:
\begin{empheq}[left=\empheqlbrace]{align}
&\omega_1=\omega_2+\omega_3 \label{eq:1} \\
&\vec{k}_1=\vec{k}_2+\vec{k}_3.  \label{eq:2}
\end{empheq}

\noindent  Random-phase three-wave interactions are then considered the main mechanism of the spectral energy transfer in capillary wave turbulence from low to high $k$ and $f$.

The theory \cite{Zakharov_1967} predicts the wave number spectra of the surface elevation $E_k \propto k^{-19/4}$, while the frequency spectrum is given by
\begin{equation}
\label{f_spectrum}
E_{\omega} \sim P^{1/2} \rho^{-2/3}\alpha^{1/6}\omega^{-17/6}.
\end{equation}

The WTT has been tested in several experiments \cite{Henry_EPL_2000, Wright_PRL_1996,Brazhnikov_EPL_2002,  Falcon_PRL_2007, Boudaoud_PRL_2008, Mordant_PRL_2008, van_de_Water_2009} which gave contradicting results. Some papers report good agreement of the spectrum shape with theory \cite{Henry_EPL_2000, Wright_PRL_1996, Brazhnikov_EPL_2002, Falcon_PRL_2007}, while some others report deviations from the expected $f$ or $k$ power-law scalings \cite{ Boudaoud_PRL_2008, Mordant_PRL_2008, van_de_Water_2009}.

Since in many laboratory experiments surface waves are excited parametrically, it is important to clarify whether the assumptions of the WTT are satisfied there. In particular, the assumptions are: (a) randomness of the waves phases, (b) infinite-box size, (c) the dominance of the three-wave interactions as the main nonlinear wave process. It is important to recall that the frequency spectra observed in experiments with the monochromatic parametric forcing of capillary waves are often dominated by multiple harmonics of the first subharmonic of the forcing frequency (e.g. \cite{van_de_Water_2009,Punzmann_PRL_09}). These harmonics are observed starting from the threshold of the parametric excitation. At higher drive the spectra may broaden showing the power-law continuum \cite{Wright_PRL_1996,Henry_EPL_2000,Brazhnikov_EPL_2002,Punzmann_PRL_09}. It is not understood (a) how the harmonics are generated and (b) how the power-law continuum is formed. Understanding the above effects and assumptions is necessary for evaluation of (dis)agreement between experiments and theory.

Another important question which needs to be addressed with regard to turbulence formation in parametrically driven capillary waves is what type of the wave-wave interactions is dominant in each of the regimes (low/high excitation drive). It is usually assumed that the three-wave interactions is the only dominant process here. Parametric excitation can be viewed as a degenerate process of a three-wave decay of the infinitely long wave, which has a wave number $\vec{k}_0 = 0$ and a frequency $\omega_0$, into two oppositely propagating subharmonic waves:
\begin{empheq}[left=\empheqlbrace]{align}
&\omega_0=\omega_1+\omega_2=2\omega_1 \label{eq:5} \\
&\vec{k}_0=\vec{k}_1+\vec{k}_2 \approx 0,  \label{eq:6}
\end{empheq}
\noindent where $\vec{k}_1=-\vec{k}_2$ and $\omega_1=\omega_2=(\alpha k_1^3/\rho)^{1/2}$.
This leads to the generation of the waves at the frequency $\omega_1=\omega_0/2$ (Faraday waves).

Apart from the decay of the $\vec{k}_0 = 0$ wave, simultaneous satisfaction of the Eqs.\eqref{eq:1} and \eqref{eq:2} is severely restricted for capillary (and capillary-gravity) waves, see e.g. \cite{McGoldrick_65, Kartashova_2010}. The selection rules favor interacting triads with disparate wave vectors, for example, those with $k_1<<k_2,k_3$, and the angles between interacting wave vectors in a narrow band between $75\,^{\circ}$ and $82\,^{\circ}$. In the presence of discrete spectra which contain only narrow multiple frequency harmonics, as is commonly observed, the above restriction practically forbids three-wave interactions unless spectral lines sufficiently broaden. However, before the broad spectra are formed, no three-wave processes are allowed, except for the parametric decay of the forcing.

In this Letter we summarize recent findings and present new results on the formation of continuous frequency spectra of capillary waves during parametric excitation. At modest damping (in distilled water) parametrically excited waves are unstable to small perturbations of the wave amplitude. This modulation instability seems to be responsible for the randomization of the wave phases, the detachment of the wave field from the container walls, spectral broadening, and gradual transition from discrete to continuous spectra.

The experimental setup is similar to that described in \cite{Punzmann_PRL_09,Shats_PRL_10}. Waves are excited parametrically in a vertically shaken container. Circular containers of the inner diameters $>$ 15 cm filled with distilled water to the heights of $\geq$ 20 mm are mounted on the top plate of the $4 kN$ electrodynamic shaker. The container is shaken at the frequency of $f_0=$(0-3500) Hz. The acceleration $a$ can be varied in a broad range. The amplitude of the shaker vibrations $\Delta_0$ is proportional to the acceleration at a given oscillation frequency, $\Delta_0 \propto a$. Below we will refer to the acceleration above the threshold of parametric excitation $\Delta a = a - a_{th}$, rather than to its absolute value.

The surface elevation of the fluid is measured using both the light diffusion and the laser reflection techniques described in \cite{Punzmann_PRL_09,Shats_PRL_10}. In the light diffusion technique, skim milk is added to the water as a diffusing agent as proposed in \cite{Henry_EPL_2000}. A thin laser beam is shined vertically from below the cell. A diffusive screen is mounted above and near the surface to collect the transmitted light (whose intensity is proportional to the surface height), which is imaged into a photo-multiplier tube. The transmission of a thin laser beam reduces the problem of the surface averaging (the "cross-over length" effect discussed in \cite{Henry_EPL_2000}). In our setup the diameter of the probing laser beam at the fluid surface ($d_B \approx 0.2$ mm ) is substantially smaller than typical wavelengths of capillary waves in the frequency range below 2 kHz ($\lambda \sim 0.5$mm) such that the measured spectra are not affected by the spatial averaging and the intensity of the transmitted light remains proportional to the wave height $\eta$. As a result, measurements can be compared directly with the theory predictions. At low wave amplitudes (e.g. at the high frequency excitation) we use a more sensitive technique based on the reflection of a thin laser beam ($< 0.5$ mm). This technique is not used at high forcing levels to avoid signal clipping by steep waves. Spectra obtained using data from both techniques are very similar.

\begin{figure}
\onefigure[width=7.0cm]{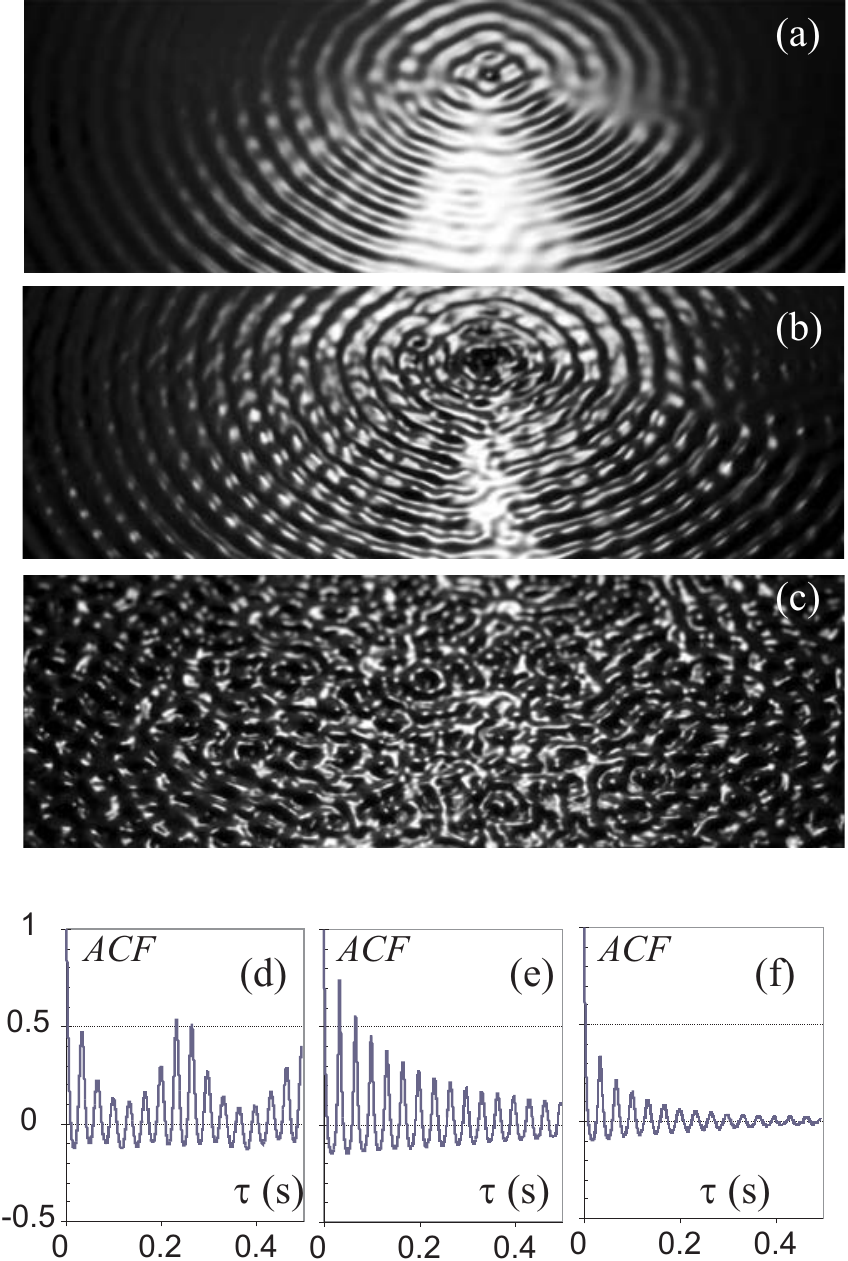}
\caption{Snapshots of the wave field evolution during the startup:(a) $t$=0.125 s, (b) $t$=0.25 s, (c) $t$=0.625 s. Autocorrelation functions of the surface elevation during parametric excitation at $f_0 = 60$ Hz at the accelerations of (d) $\Delta a = 0.25g$, (e) $\Delta a = 0.6g$ , (f) $\Delta a = 1.3g$ .}
\label{fig1}
\end{figure}

First, we consider the dynamics of the wave excitation captured using fast video camera. Figures~\ref{fig1}(a-c) show three snapshots of the wave field filmed at the intermediate acceleration in a cylindrical container shaken vertically at the frequency of 250 Hz. Initially, 0.125 s after the forcing is switched on, Fig.~\ref{fig1}(a), the wave fronts of the excited wave are concentric circles, reflecting the shape of the wall. However very soon these wave fronts become azimuthally modulated, Fig.~\ref{fig1}(b), and after about 0.6 s the wave field appears to be severely disordered, as in Fig.~\ref{fig1}(c). Similar evolutions are observed for all excitation frequencies.

Order-disorder transitions in capillary waves have been studied in \cite{Tufillaro_PRL_89}, where it has been shown than the radial correlation length is dramatically reduced at rather low excitation levels. When the transition occurs, the bulk wave field is no longer connected to the wall since the correlation length becomes much shorter than the container radius. At this point finite-box size effects become unimportant. Similarly, we observe a dramatic change in the autocorrelation time with the increase in the acceleration. Figures~\ref{fig1}(d-f) show the autocorrelation functions of the surface elevation measured at the mid-radius at three levels of accelerations. At low acceleration, the autocorrelation function shows two characteristic frequencies: 30Hz of the first subharmonic of the excitation frequency, and a modulation at about 4-5 Hz corresponding to the eigenmode of the cylindrical container (wave length of $~ 14$ cm, which is the container radius), Fig.~\ref{fig1}(d). At higher forcing, the autocorrelation function becomes exponentially decaying, as in Fig.~\ref{fig1}(e). Further increase in the acceleration leads to a fast decrease in the auto correlation time down to 1-2 periods of the first subharmonic wave (30 Hz) illustrated in Fig.~\ref{fig1}(f).

Such a transition from order to disorder is due to the development of the modulation instability of capillary waves. This instability, which was first considered theoretically by Zakharov in 1968 \cite{Zakharov1968}, is the capillary counterpart of the well-known Benjamin-Feir instability of gravity waves \cite{Benjamin&Feir67}. Modulation instability (MI) of capillary waves has been identified for the first time in \cite{Punzmann_PRL_09}. It develops when the Lighthill criterion, $(\partial \omega/\partial|a|^2)(\partial^2 \omega_k/\partial k^2)<0$, (where $\omega= \omega_k [1-(ka)^2/16]$ includes a nonlinear frequency correction \cite{Debnath_94}) is satisfied. The development of the MI is observed at a given point in space as the modulation of the envelope which, at higher wave amplitudes (forcing) leads to  breaking of the initially continuous waves into sequence of the envelope solitons \cite{Punzmann_PRL_09}.

\begin{figure}
\onefigure[width=6.5cm]{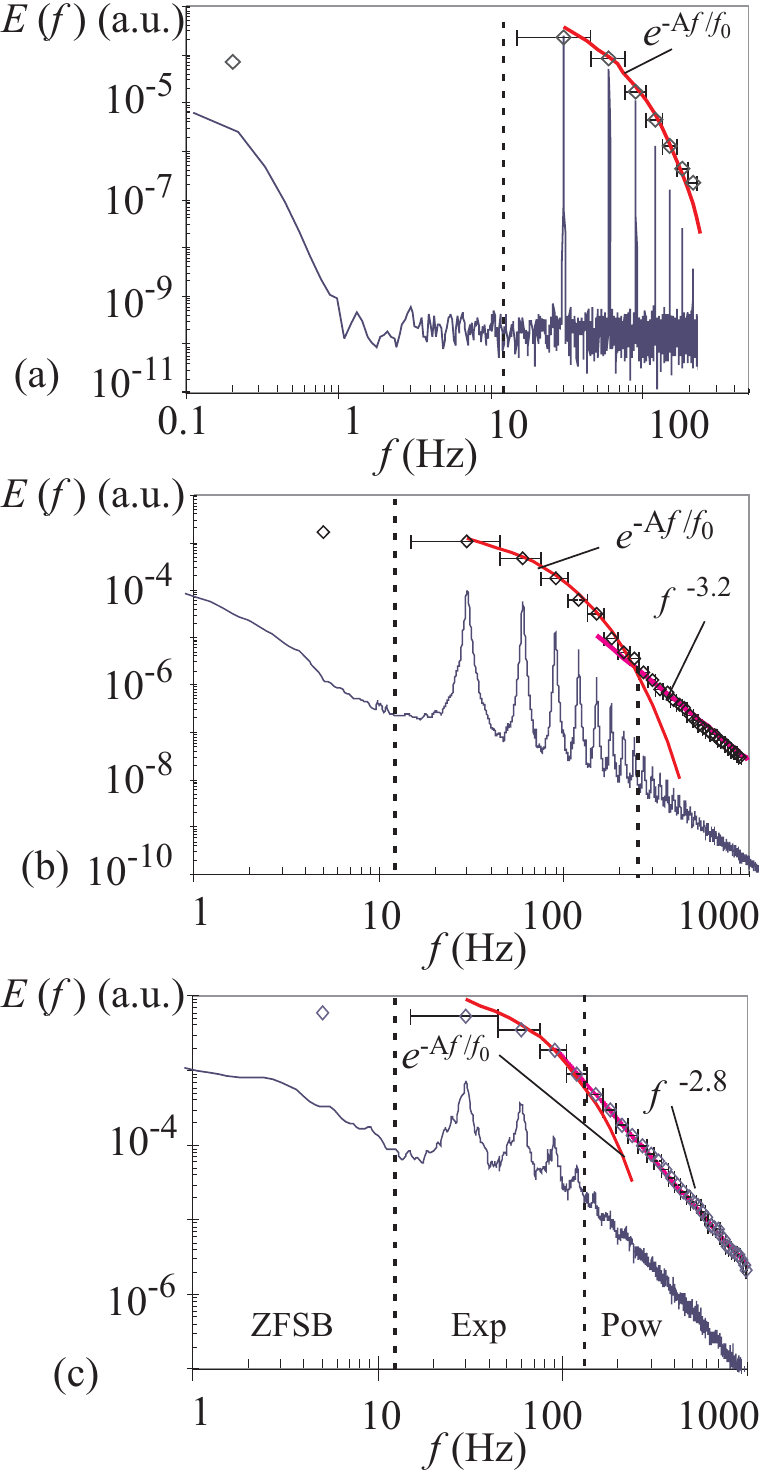}
\caption{ Frequency spectra of capillary waves at different accelerations. (a)  $\Delta a = 0.5g$, (b) $\Delta a = 1.4 g$ and (c) $\Delta a = 2.1 g$. Open diamonds show the spectral powers $E_f$ of each harmonics.}
\label{fig2}
\end{figure}

To understand how and when the transition to the power-law continuum occurs, we now consider frequency spectra measured at various levels of acceleration. Figure~\ref{fig2} shows frequency spectra measured at three different accelerations, up to the threshold of the droplet generation. At the lowest acceleration, just above the parametric excitation threshold, we observe several (up to 10) frequency harmonics $f_n=nf_1$ which become broader as the acceleration is increased. To obtain a spectral power contained in each of the harmonics, spectral densities $E(f)$ are integrated within the frequency bands $[f_n-f_1/2,f_n+f_1/2]$. The resulting spectral powers $E_f$ are shown by open diamonds above the corresponding spectral peaks in Figs.~\ref{fig2}(a-c).

At low forcing ($\Delta a=0.5 g$), Fig.~\ref{fig2}(a), the harmonic amplitudes exponentially decay as a function of the wave harmonic, $E(nf_1) \sim e^{-n}$. The band-integrated amplitudes of the harmonics in Fig.~\ref{fig2}(a) also follow the geometric progression law $E_f=E_1e^{-Af/f_1}$, where $A=1.2$. Along with the harmonics, a zero-frequency sideband (ZFSB) is observed at all levels of forcing, even at the lowest one ($f<1$ Hz). Both the harmonics and the ZFSB are observed just above the threshold of the parametric excitation of the first harmonic, at $\Delta a \gtrsim 0$. The ZFSB results from the modulation instability of capillary waves.

The generation of the multiple frequency harmonics, however is not really understood. Linear analysis of the stability of the interface between two fluids \cite{Kumar_94} shows that the parametric excitation leads to the instability zones corresponding to the multiple frequency harmonics and, in principle, could explain the harmonic generation. However, the acceleration thresholds for the generation of harmonics with $n>4$ expected for water are much higher ($20-30g$ at $f_0=50 Hz$) \cite{Bechhoefer_95} than observed in our experiments ($0.2-0.5g$ at $f_0=60 Hz$). Another possible interpretation is related to the WTT energy flux, which, as was shown in numerical simulations \cite{Falkovich_Shafarenko_1988} can lead to the accumulation of energy in the harmonics of the forcing frequency. This however is unlikely to be the case in this experiment since the harmonics are observed at very low forcing levels, when all wave harmonics are perfectly coherent \cite{Punzmann_PRL_09}, which is inconsistent with the assumption of the random phases.

An intermediate regime is achieved at the forcing level of $\Delta a = 1.4g$. Fig.~\ref{fig2}(b) shows the spectrum in this case. The spectrum is also dominated by discrete harmonics, however spectral broadening of the harmonics leads to their overlapping and to the formation of spectral continuum. It can be seen that up to the frequencies of about 220 Hz, the spectral power $E_f$ follows an exponential fit, $E_f=E_1e^{-Af/f_1}$ ($A=0.9$), while at higher frequencies the spectrum scales according to a power law, $E_f \sim f^{-3.2}$. The ZFSB in this regime extends to about 10 Hz.

At the highest level of forcing ($\Delta a=2.1 g$), Fig.~\ref{fig2}(c), the spectrum changes further, showing exponential range $E_f=E_1e^{-Af/f_1}$ ($A=0.8$) for the first 4 harmonics and an extended continuum which is well fitted by a $f^{-2.8}$ power law for the frequencies above 120 Hz. The $E(f) \sim f^{-2.8}$ part of the spectrum, which extends over one decade, is in a good agreement with the WTT $E \sim \omega^{-17/6}$ prediction, Eq.~(\ref{f_spectrum}). The energy contained in the ZFSB in this regime is the highest.

\begin{table}
\centering 
\begin{tabular}{c c c c c} 
\hline\hline 
Figure & $\Delta a (g)$ & $E_{ZF}/E_{T}$  & $E_{Exp}/E_{T}$ & $E_{Pow}/E_{T}$ \\ [0.5ex] 
\hline 
2(a) & 0.5   & 0.18 & 0.82 & 0.0  \\ 
2(b) & 1.4   & 0.48 & 0.51 & 0.01 \\
2(c) & 2.1   & 0.31 & 0.46 & 0.23 \\ [1ex] 
\hline 
\end{tabular}
\label{table:damping} 
\caption{Energy distribution in various spectral bands corresponding to spectra of Fig.~\ref{fig2}. Energy is shown as a fraction of the total energy $E_T=\int_{0}^{2kHz} E(f)df$. $E_{ZF}$, $E_{Exp}$ and $E_{Pow}$ refer to the zero-frequency sideband, exponential, and the power-law ranges respectively. } 
\end{table}

Table 1 shows the energy distribution between three spectral bands identified above: ZFSB, exponential (discrete harmonics) and the power-law (continuum) ranges. The forcing injects energy into the exponential (discrete harmonic) spectral range, which is then redistributed into other ranges via wave-wave interactions. At low levels of forcing ($\Delta a=0.5 g$) no transfer of energy into the high-frequency  continuum is observed, while 18\% of the total spectral energy is coupled into the ZFSB. At the intermediate forcing, corresponding to Fig.~\ref{fig2}(b), only 1\% of total energy is contained in the power-law continuum, while the rest is distributed almost equally between ZFSB and the exponential range. Finally, at the highest forcing, 23\% of energy is delivered from the forcing range into the power-law continuum, leaving 31\% in the ZFSB.

This can be explained by recalling that at low forcing no three-wave interactions between discrete harmonics are possible. As the drive is increased, the satisfaction of the three-wave matching rules of Eqs.\eqref{eq:1} and \eqref{eq:2} in this situation becomes possible due to the spectral broadening of the harmonics, when broadening compensates for the frequency mismatch $\Delta\Omega$. This broadening is due to the modulation instability, a four-wave process of the sideband generation. Thus, a four-wave process leads to the nonlinear broadening of the harmonics and to the generation of the ZFSB until broadening exceeds $\Delta\Omega$ such that three-wave interactions are allowed. This opens the gate for the energy flux into the power-law continuum. The power law tail $E_f \sim f^{-2.8}$, observed at the highest forcing, is in agreement with the the WTT scaling of $\omega^{-17/6}$. The random-phase assumption is justified at high forcing as was reported in \cite{Punzmann_PRL_09}. Also, the build-up of the spectral continuum increases the range of waves satisfying Eqs.\eqref{eq:1} and \eqref{eq:2} and opens the way for the three-wave interactions.

The evidence in support of the four-wave interaction process in the harmonic-dominated regime has been presented in \cite{Shats_PRL_10}. If the wave harmonics are modulationally unstable, each of them satisfies four-wave synchronism conditions: $\omega_{n-}+\omega_{n+}=2\omega_n$, $k_1+k_2=2k_n$, or in another form, $\omega_{n-, n+}=\omega_n \pm \Omega$, $k_{1,2}=k_n \pm K$, where $\Omega$ is the modulation frequency \cite{Zakharov_Ostrovsky_2009}. In this case the wave harmonics $\omega_n$ are all phase coupled via $\Omega$, represented in Fig.~\ref{fig2} by the ZFSB range. The coherent-phase \textit{four-wave} coupling has indeed been found in this system. As has been reported in \cite{Shats_PRL_10}, the maximum tricoherence, which is a degree of a coherent four-wave coupling, is high in the conditions corresponding to the discrete spectra of capillary waves. This result also indicates that the four-wave process (cubic nonlinearity) is at work in the conditions when the three-wave interactions (quadratic nonlinearity) is forbidden by the Eqs.\eqref{eq:1} and \eqref{eq:2}.

\begin{figure}
\onefigure[width=6.5cm]{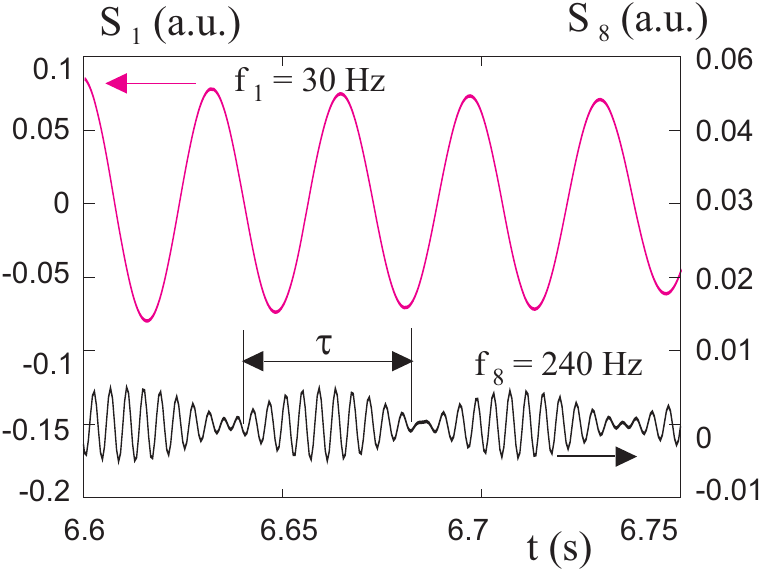}
\caption{Time evolution of the $n=1$ frequency harmonic, $f=(30\pm15) Hz$, (top) and of the $n=8$ harmonic at $f=(240\pm15) Hz$ for the intermediate forcing regime of Fig.~\ref{fig2}(b).}
\label{fig3}
\end{figure}

Let us now analyze physical conditions which determine the transition of the exponential part of the spectrum into the power-law continuum. We denote the cross-over frequency where these two ranges meet as $f_{co}$. It can be seen from Figs.~\ref{fig2}(b,c) that $f_{co}$ decreases with the increase in forcing from $f_{co} \approx 220$ Hz at $\Delta a=1.4 g$ to $f_{co} \approx 120$ Hz at $\Delta a=2.1 g$. We have previously reported that the spectral line shapes of the broadened wave harmonics can be accurately represented using the hyperbolic secant \cite{Punzmann_PRL_09}: $E_n \propto sech [b_n(f-f_n)]$. In the time domain such spectra correspond to the strongly modulated sinusoidal pulses $s_n(t) = (\pi/b_n)sech [\pi^2/(b_nt)]e^{if_nt}$ which are very similar to the well known envelope solitons. Statistically averaged lifetime $\tau_n$ (length) of such a pulse is related to the spectral width parameter $b_n$ as $\tau_n=b_n/\pi^2$. The loss of coherence in the wave-wave interactions occurs when the lifetime $\tau_n$ of a highest-frequency interacting wave becomes shorter than the period of the first harmonic $T_1$: at $\tau_n \lesssim T_1$. We find that exactly this condition is satisfied at $f=f_{co}$: $\tau_{co} \approx T_1$. Figure~\ref{fig3} shows time evolution of the two wave harmonics bandpass-filtered from the frequency bands $[f_n-f_1/2,f_n+f_1/2]$, for $f_1=30$ Hz ($n=1$) and $f_8=240$ Hz ($n=8$) for the data corresponding to Fig.~\ref{fig2}(b). The second frequency is chosen such that $f_8 \gtrsim f_{co} \approx 220$ Hz. One can see that the average length of the $f_8$ pulse, $\tau$, is indeed comparable to the period of the first harmonic $T_1=1/f_1$. This result is confirmed by calculating $b_n$ from the spectra at $f_n \gtrsim f_{co}$ at several levels of forcing and by comparing $\tau_n=b_n/\pi^2$ with $T_1$.

Naturally, when $\tau_{n} <  T_1$, the nonlinear interaction time between the 1st and the $n$-th harmonics decreases dramatically turning coherent interaction into the random-phase one. Thus, we conclude here that the criterion corresponding to the transition from the exponential to the power-law range coincides with the criterion of phase-randomization in the wave-wave interactions, namely, $\tau/T_1 < 1$.

\begin{figure}
\onefigure[width=5.0cm]{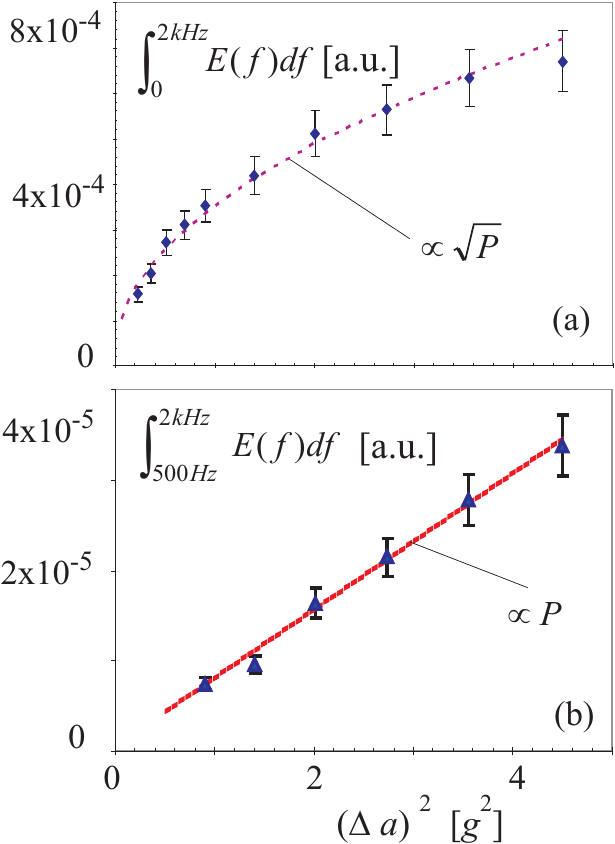}
\caption{Spectral energy of capillary waves contained in different frequency ranges versus the energy flux injected by forcing ($ P \propto \Delta a^2$): (a) for the entire spectral range from 0 to 2kHz and (b) in the power-law range, from 500Hz to 2kHz.}
\label{fig4}
\end{figure}

The observations of the power-law scaling and of the random phases above $f>f_{co}$ validate the comparison of the continuum spectral range with the WTT. We test the scaling of the spectral energy with the energy flux $P$ given by Eq.~\eqref{f_spectrum}. In our experiments the energy input is proportional to the square of the wave amplitude, which is proportional to the square of the acceleration above the threshold of the parametric instability,  $P \propto \Delta a^2$. Figure~\ref{fig4}(a) shows the total energy in the spectrum as a function of $\Delta a^2$. The spectral power integrated from 0 to 2 kHz varies as $\sqrt{P}$. This integral is however dominated by the exponential and by the ZFSB ranges. To test the scaling of the spectral energy in the power-law range we compute $\int_{0.5kHz}^{2kHz}{E(f)df}$. Fig.~\ref{fig4}(b) shows that this integral scales linearly with $\Delta a^2$, contradicting the $E_{\omega} \sim P^{1/2}$ prediction of the WTT. A similar observation of the linear $E_f$ \textit{vs} $P$ dependence was already reported in experiments on surface waves by Falcon \textit{et al.} \cite{Falcon_PRL_2007}, which cast a doubt on the agreement between experiments and the WTT. In that experiment waves were excited at rather low frequency and the spectra showed the power-law scalings over the entire spectral range.

\begin{figure}
\onefigure[width=5.0cm]{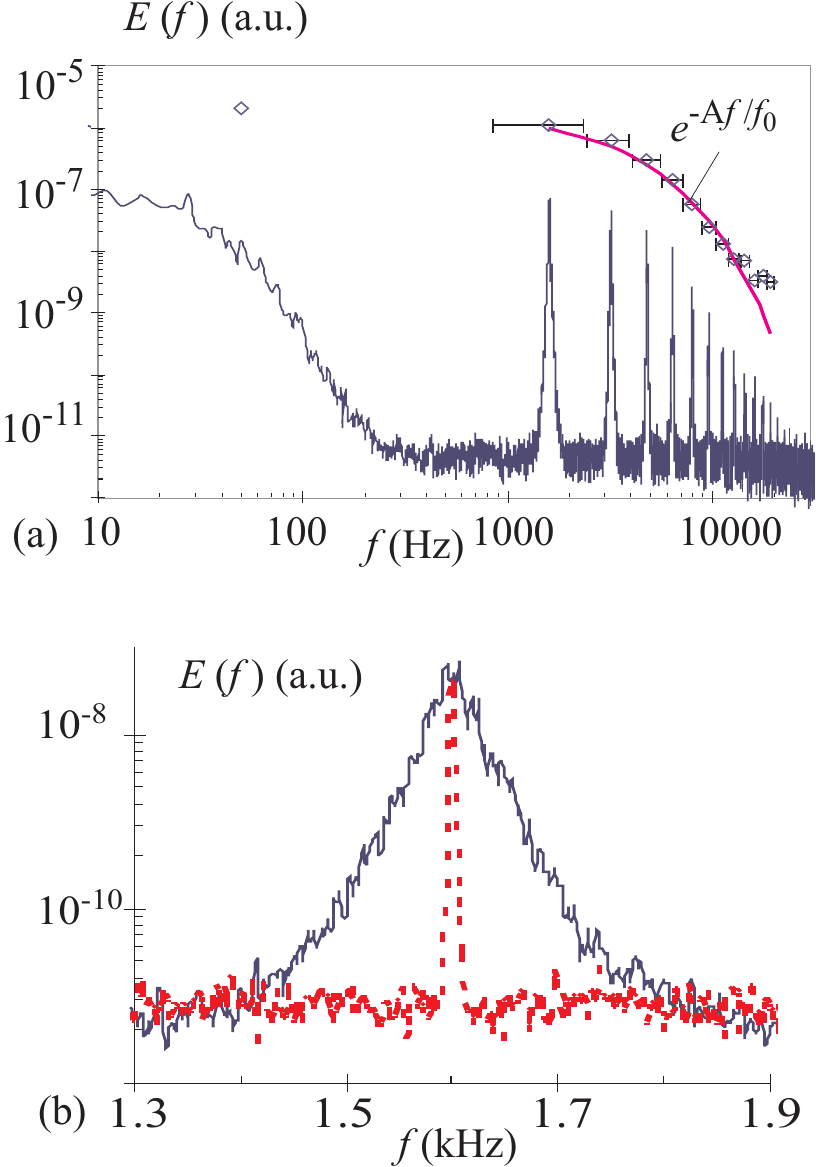}
\caption{(a) Spectrum in the regime of high-frequency excitation at $f_0=3.2$ kHz, high forcing at $\Delta a=65 g$. A constant in the exponential fit $A=$0.7. (b) Zoomed-in part of the same spectrum. Dashed red spectral line corresponds to low forcing at $\Delta a=3 g$.}
\label{fig5}
\end{figure}

Finally, to illustrate that our results do not depend on the container size, we performed experiments at much higher excitation frequency, $f_0=$3.2 kHz. In this case, a circular container of the diameter $d=$180 mm accommodates 160 wave lengths on the radius (first harmonic, $f_1=$1.6 kHz, $\lambda \approx 0.56$ mm), substantially more than at $f_1=$30 Hz shown in Fig.~\ref{fig1}. The spectrum for the case of the high-frequency excitation is shown in Fig.~\ref{fig5}. It has all the features observed at low frequency: discrete spectrally broadened peaks, exponential fit for the integrated spectral energy, and the ZFSB. Note that the spectral width of the ZFSB extends up to 200 Hz at high drive. Broadening of the harmonics is quite similar to that at low frequency, showing characteristic exponential tails (triangular line shapes on the log-linear scale), Fig.~\ref{fig5}(b). Thus, we conclude that the finite-box size effects are not important in these experiments above the threshold of modulation instability.

Summarizing, we have shown that modulation instability plays a major role in the transition from monochromatic parametrically excited wave to the broadband spectra. The instability is responsible for the strong modulation of the wave amplitude and the breaking of the continuous waves into envelope solitons. This leads to a substantial shortening of the autocorrelation time and the correlation length which results in the detachment of the wave field from the container walls. At the same time, this very process leads to the spectral broadening of the wave harmonics which start overlapping to generate the wave continuum. In the continuum three-wave interactions between two high-frequency waves and one low-frequency ($f_1$) harmonic become possible. Shorter correlation times of the wave harmonics due to the MI randomize phases in the wave-wave interactions, leading to the conditions assumed in the WTT. When these conditions are satisfied, a power-law continuum is observed.

Three distinct ranges in the spectra of the parametrically driven capillary waves have been identified. Spectral energy contained in multiple harmonics obeys the exponential fit $E_f=E_1e^{-Af/f_1}$ with $A$ slowly decreasing with the increase in forcing from $A=1.2$ to $A=0.8$. Wave harmonics spectrally broaden as the forcing is increased. This broadening is observed for the first subharmonic and for its harmonics very close to the threshold of parametric instability. The broadening increases proportionally to the forcing (acceleration), while the spectral line shapes remain unchanged, $E_n \propto sech [b_n(f-f_n)]$.

Above some critical broadening of the wave harmonics, a continuum develops in the spectrum, starting at higher frequencies. This continuum is characterized by a power-law scaling, which at the highest drive shows $E_f \sim f^{-2.8}$, close to the one predicted by the WTT. However, the prediction of the dependence of the spectral energy on the injected energy flux differs from the theoretically predicted $E_f \propto P^{1/2}$: in experiments $E_f$ scales linearly with the injected power $P$. Such a discrepancy may be due to the fact that the energy delivered into the power-law continuum is dissipated there, rather than being cascaded further down the spectrum. In other words, the assumption of the inertial interval in the continuum is not satisfied.

We have found a criterion for the transition from the exponential scaling of $E_f$ to the power law scaling. This occurs at the frequency of the wave harmonics whose average length (time) of the modulation envelope becomes shorter than the period of the first harmonic, at $\tau_{n} <  T_1$. This criterion corresponds to a strong reduction in the nonlinear interaction time between the first and the $n$-th wave harmonics, or in other words, leads to the phase-randomization in wave-wave interactions.

Results obtained in this paper are not dependent on the size of the container: this has been checked by varying the frequency (and wave length $\lambda$) of the first harmonic by a factor of more than 50 (16 for $\lambda$) at the given size of the container.

\acknowledgments
The authors thank G. Falkovich and E. Kartashova for many valuable comments. This work was supported by the Australian Research Council's Discovery Projects funding scheme (DP0881571).

\end{document}